\documentclass[a4paper,11pt,superscriptaddress]{article}
\usepackage{graphicx}
\usepackage{dcolumn}
\usepackage{bm}
\usepackage[latin1]{inputenc}
\usepackage{fancybox}
\usepackage{float}
\usepackage{hhline}
\begin{document}
\title{Acoustic wave propagation in a macroscopically inhomogeneous porous medium saturated by a fluid}
\author{L. De Ryck\thanks{Laboratorium voor Akoestiek en Thermische Fysica, K. U. Leuven, Celestijnenlaan 200D, 3001 Heverlee, Belgium.}, J.-P. Groby\thanks{Laboratorium voor Akoestiek en Thermische Fysica, K. U. Leuven, Celestijnenlaan 200D, 3001 Heverlee, Belgium.}, P. Leclaire\thanks{Laboratorium voor Akoestiek en Thermische Fysica, K. U. Leuven, Celestijnenlaan 200D, 3001 Heverlee, Belgium.}, W. Lauriks\thanks{Laboratorium voor Akoestiek en Thermische Fysica, K. U. Leuven, Celestijnenlaan 200D, 3001 Heverlee, Belgium.},\\ A. Wirgin\thanks{Laboratoire de M\'ecanique et d'Acoustique, UPR 7051 CNRS, 31 chemin Joseph Aiguier, 13009 Marseille, France.}, C. Depollier\thanks{Laboratoire d'Acoustique de l'Universit¨\'e du Maine, Avenue Olivier Messiaen, 72000 Le Mans, France.}, and Z. E. A. Fellah\thanks{Laboratoire de M\'ecanique et d'Acoustique, UPR 7051 CNRS, 31 chemin Joseph Aiguier, 13009 Marseille, France.}}
\maketitle
\begin{abstract}
The equations of motion in a macroscopically inhomogeneous porous medium saturated by a fluid are derived. As a first verification of the validity of these equations, a two-layer rigid frame porous system considered as one single porous layer with a sudden change in physical properties is studied. The wave equation is derived and solved for this system. The reflection and transmission coefficients are calculated numerically using a wave splitting-Green's function approach (WS-GF). The reflected and transmitted wave time histories are also simulated. Experimental results obtained for materials saturated by air are compared to the results given by this approach and to those of the classical transfer matrix method (TMM).
\end{abstract}
{\emph{Introduction.---}}An inhomogeneous medium is one with properties that vary with position. Inhomogeneous and layered materials can be found in many fields of physics, for instance in optics and electromagnetism \cite{Aspnes,Osamu,PIER} or in acoustics \cite{Wirgin,PRL94}. Other examples are geophysical media \cite{Berryman}, granular \cite{TournatPRL,Jia} or porous materials with depth-varying physical properties. The study of the acoustic wave propagation in inhomogeneous porous and granular media is of great interest in building engineering, in petroleum prospection and civil engineering.

Acoustic wave propagation in fluid-saturated porous media is relatively well-known thanks to the early work of Biot \cite{Biot56} and to the contribution of many researchers \cite{John87,Att1,Allard} since then. A porous medium can be defined as a biphasic material with a solid phase and a fluid phase. This definition encompasses sandstones, concrete, absorbing polyurethane foams, snow or bones for example. While homogenized porous media have been extensively studied, macroscopically-inhomogeneous porous media have received far less attention.

In this article, acoustic propagation in fluid-saturated macroscopically-inhomogeneous porous materials is studied. It is assumed that the wavelengths are greater than the average heterogeneity size at the pore scale so that the physical properties of the material are homogenized. However, the homogenized properties can vary with the observation point within the material at the macroscopic scale of the specimen. The equations of motion are derived from Biot's alternative formulation of 1962 \cite{BiotJAP62} in which the total stress tensor, the fluid pressure, the solid displacement and the fluid/solid relative displacement $\textbf{w}$ are used. It was briefly stated by Biot \cite{BiotJAP62} and confirmed \cite{Carcione} that these variables should be employed to describe the acoustical properties of porous media with inhomogeneous properties. Among many possible applications, this work is a first contribution towards the determination of the inhomogeneity profile of unknown materials by the use of inversion methods.

The equations of motion for a macroscopically inhomogeneous medium with elastic skeletton are derived. A first verification of the validity of the proposed equations is the study of a porous material saturated by air in the rigid frame approximation \cite{Allard}, i.e. when the fluid is light and the solid skeleton therefore relatively immobile. The porous material is then considered as an equivalent fluid with frequency-dependent and depth-dependent effective density and bulk modulus. In this case, a wave equation is derived and solved numerically for a two-layer porous system treated as one single porous medium with a sudden but continuous change in physical properties. This provides an excellent means of comparing the proposed method (the WS-GF method) to the results of the well established Transfer Matrix Method (TMM) developed to calculate the acoustical properties of multilayer porous systems \cite{Allard}. The WS-GF method is applicable to any depth-dependent inhomogeneous system and the two-layer system is just chosen for a testing purpose. An inhomogeneity function with a shape similar to the Heaviside function, but which is analytical, is used. The jump with controlable steepness is created by multiplying the parameters by the inhomogeneity function. The thickness of the inhomogeneous medium is equal to the sum of thicknesses of each layer.

The reflected and transmitted pressure fields are calculated from a known incident plane wave impinging at normal incidence using a Wave Splitting-Green's functions approach (WS-GF) \cite{PIER}. The method can be extended to include oblique incidence by working on the appropriate component of the wavevectors. The reflection and transmission coefficients and the time histories of the reflected and transmitted waves simulated with this technique are compared to the results given by the classical transfer matrix method (TMM) for multilayer porous materials \cite{Allard}. Experimental results obtained in a two-layer material saturated by air are compared to the simulations.

{\emph{The equations of motion in an inhomogeneous poroelastic medium.---}}The constitutive linear stress-strain relations in an initially stress-free, statistically-isotropic porous medium can be written as \cite{BiotJAP62} 
\begin{eqnarray}
 \sigma_{ij} & = & 2\mu \epsilon_{ij} + \delta_{ij}(\lambda_{c} \theta - \alpha M \zeta), \label{sigma} \\
 p           & = & M\left(-\alpha \theta + \zeta \right), \label{p}
\end{eqnarray}
\noindent where $\sigma_{ij}$ is the total stress tensor and $p$ the fluid pressure in the pores; $\delta_{ij}$ denotes the Kronecker symbol (the summation on repeated indices is implied); $\theta = \nabla \cdot \mathbf{u}$ and $\zeta = -\nabla \cdot \mathbf{w}$ are respectively the dilatation of the solid and the variation of fluid content where $\mathbf{u}$ is the solid displacement and $\mathbf{w} = \phi \left(\mathbf{U} - \mathbf{u} \right)$ the fluid/solid relative displacement ($\mathbf{U}$ is the fluid displacement); $\phi$ is the porosity; $\epsilon_{ij} = \frac{1}{2}\left(u_{i,j} + u_{j,i}\right)$ the strain tensor of the solid (the comma denotes spatial partial derivatives); $\lambda_{c} = \lambda + \alpha^{2} M$, where $\lambda$, $\mu$, $M$ are elastic constants and $\alpha$ a coefficient of elastic coupling. These parameters were defined by Biot and Willis \cite{BiotWillis}.

Applying the momentum conservation law in the absence of body forces, the equations of motion are written
\begin{eqnarray}
\nabla \! \cdot \! \boldsymbol{\sigma} & = & \rho \mathbf{\ddot{u}} + \rho_{f}\mathbf{\ddot{w}}, \label{moment1} \\ 
- \nabla p & = & \rho_{f} \mathbf{\ddot{u}} + m \mathbf{\ddot{w}} + \frac{\eta}{\kappa} F \mathbf{\dot{w}}, \label{moment2}
\end{eqnarray}
where the dot and double dots notations refer to first and second order time derivatives, respectively; $\rho_{f}$ is the density of the fluid in the (interconnected) pores, $\rho$ the bulk density of the porous medium, such that $\rho = (1-\phi) \rho_{s} + \phi \rho_{f}$ where $\rho_{s}$ is the density of the solid; $\displaystyle m = \rho_{f} \tau_{\infty} / \phi$ is a mass parameter defined by Biot \cite{BiotJAP62}, $\tau_{\infty}$ is the tortuosity, $\eta$ the viscosity of the fluid, $\kappa$ the permeability and $F$ the viscosity correction function. This function has been studied in detail by Johnson et al. \cite{John87} and by Allard \cite{Allard}.

We shall now derive the equations of motion for an inhomogeneous porous layer or a half space whose properties vary along the depth $x$. The fact that the medium is inhomogeneous is not incompatible with the fact that it is isotropic as the inhomogeneous medium can be considered as a superposition of an infinite number of thin isotropic sub-layers of thickness $dx$. Therefore, the following parameters in the above equations are now dependent on $x$: $\lambda$, $\mu$, $\lambda_{c}$, $\alpha$, $M$, $\phi$, $\rho$, $m$, $\tau_{\infty}$, $\kappa$ and $F$. The ratio $\eta/\kappa$ is the flow resistivity and is often used instead of $\kappa$ in engineering acoustics applications. It is denoted by $R_f$ here. The viscosity correction function $F$ incorporates the viscous characteristic length $\Lambda$ of Johnson et al. \cite{John87} and the thermal characteristic length $\Lambda'$ of Champoux and Allard \cite{Champ}. These parameters also depend on $x$. Inserting equations (\ref{sigma}) and (\ref{p}) into equations (\ref{moment1}) and (\ref{moment2}) yields the equations of motion in term of the displacements
\begin{equation}
\left\{ \!
\begin{array}{l}
 \nabla \! \left[ \left( \lambda_c \! + \! 2 \mu \right) \! \nabla \! \cdot \! \mathbf{u}  + \alpha M \nabla \! \cdot \! \mathbf{w} \right]\! -\! \nabla \! \wedge \! \left[ \mu \nabla \! \wedge \! \mathbf{u} \right]\! -\! \\ [8pt] 
\! 2 \nabla \mu \nabla\! \cdot\! \mathbf{u}\!+ \!2 \nabla \mu \!\wedge \!\left( \nabla \!\wedge\! \mathbf{u}\right)\!+\!2\left[\nabla \mu \! \cdot \! \nabla \right]\mathbf{u}\!=\!\rho \ddot{\mathbf{u}}\!+\!\rho_{f}\ddot{\mathbf{w}}\!, \\[10pt]
\displaystyle \nabla \left[M \nabla\! \cdot\! \mathbf{w}+\alpha M \nabla \!\cdot\! \mathbf{u}\right]=\rho_{f}\ddot{\mathbf{u}}+m \ddot{\mathbf{w}}+\frac{\eta}{\kappa} F \dot{\mathbf{w}},
\end{array}
\right.
\end{equation}
where the $x$-dependence of the constitutive parameters has been removed to simplify the notations.

{\emph{Wave equation in a rigid frame porous medium.---}}The previous equations can be applied to porous media with an elastic frame. Under the assumption of a rigid frame, $\mathbf{u}=0$ and equations (\ref{sigma})-(\ref{moment2}) can be simplified. The porous medium can be considered as an equivalent fluid (at the scale of the wavelengths) described in the frequency domain by
\begin{eqnarray}
    -j\omega p           & = & K_e(x,\omega) \nabla. [\phi(x)\mathbf{\dot{U}}], 		\label{dpdt2}\\
    - \nabla p & = & j\omega \rho_e(x,\omega) \phi(x)\mathbf{\dot{U}},                  \label{gradp2}
\end{eqnarray}
where $\rho_e(x,\omega)$ and $K_e(x,\omega)$ are respectively the effective density and bulk modulus of the inhomogeneous equivalent fluid. Their expressions are 
\begin{eqnarray}
 \displaystyle    \rho_e(x,\omega)  =  \rho_f \frac{\tau_{\infty}(x)}{\phi(x)}\left[1 - j\frac{R_f(x) \phi(x)}{\omega\rho_f \tau_{\infty}(x)} F(x,\omega ) \right], \label{adyn} \\
 \displaystyle    K_e(x,\omega)   =  \frac{\gamma P_{0}}{\phi(x)\left(\gamma\!-\!(\gamma\!-\!1) \!\left[1\!-\!j \frac{R_f(x) \phi(x)}{B^2 \omega \rho_f\tau_{\infty}(x)}G(x,\!B^2 \omega) \right]^{-1}\right)},
\end{eqnarray}
where $\gamma$ is the specific heat ratio, $P_0$ the atmospheric pressure and $B^2$ the Prandtl number. $F(x,\omega)$ and $G(x,B^2\omega)$ are the well-defined correction functions of the Johnson-Allard model \cite{John87,Allard}. The effective density and bulk modulus of the inhomogeneous equivalent fluid are functions of the frequency-independent parameters $\phi(x)$, $\tau_{\infty}(x)$, $R_f(x) = \eta/\kappa(x)$, $\Lambda(x)$ and $\Lambda'(x)$. These are the parameters that should be multiplied by the inhomogeneity function in order to account for the change in properties.

The wave equation in $p$ is obtained by combining equations (\ref{gradp2}) and (\ref{dpdt2})
\begin{eqnarray}
\omega^{2} p +K_{e}(x,\omega)\nabla \cdot \left(\frac{1}{\rho_{e}(x,\omega)}\nabla p \right)=0.  \label{WaveEq}
\end{eqnarray}

{\emph{Wave splitting technique.---}}The second order differential operator of the wave equation in an homogeneous fluid can be factorized and this yields a system of two coupled first order differential equations
\begin{equation}
	\left[\partial_x \mp \frac{j\omega}{c_0}\right]p = \pm Const \times \frac{j\omega}{c_0}p^{\pm}
 \end{equation}
where $c_0$ is the sound speed in air, $p^+$ corresponds to right-going waves and $p^-$ corresponds to left-going waves, the sum of which equals $p$. This is the so-called Wave Splitting description, which was mainly used in scattering problems in the time domain in electromagnetism \cite{Sail} and then adapted to the frequency domain \cite{PIER}. It can be seen as a change of variables from $(p, \partial_xp)$ to $(p^+,p^-)$.
\begin{figure}[ht]
	\centering
		\includegraphics[width=0.7\textwidth]{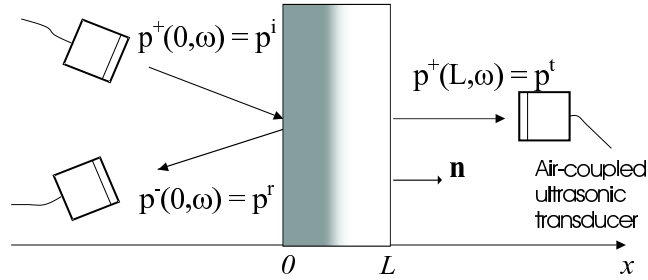}
	\caption{\emph{Slab of inhomogeneous porous material.}}
	\label{fig:slab}
\end{figure}

An inhomogeneous porous slab on which impinges an incident wave is shown in Fig. \ref{fig:slab}. Applied to the wave equation (\ref{WaveEq}), the wave splitting transformation is
\begin{equation}
	p^{\pm}  = \frac{1}{2}\left[p \pm Z_0 \phi(x)\mathbf{\dot{U}}.\mathbf{n} \right]
	\label{WS}
\end{equation}
where $Z_0=\rho_f c_0$ is the characteristic impedance of the fluid surrounding the slab and $\mathbf{n}$ the unit normal vector (Fig. \ref{fig:slab}). A system of linear first order coupled differential equation is obtained by combining equation (\ref{WS}) with equations (\ref{gradp2}) and (\ref{dpdt2}):
\begin{equation}
	\left\{
	\begin{array}{rcl}
		\partial_x p^+ & = & \phantom{-}A^+(x,\omega)p^+ + A^-(x,\omega)p^-, \\
		\partial_x p^- & = & -A^-(x,\omega)p^+ - A^+(x,\omega)p^- 
	\end{array}
	\right.
	\label{p+p-}
\end{equation}
with 
\begin{equation}
	A^{\pm}(x,\omega) = \frac{j\omega}{2}\left[\frac{Z_0}{K_e(x,\omega)} \pm \frac{\rho_e(x,\omega)}{Z_0}  \right].
\end{equation}

{\emph{Numerical resolution of the wave equation.---}}The computation principle is the following: $p^{\pm}$ are first calculated in the surrounding homogeneous fluid at $x=L$. An infinitely thin homogeneous layer of thickness $dx$ is then inserted at $x=L-dx$. The characteristic impedance of this layer is the same as the characteristic impedance of the inhomogeneous material at that point. At $x=L$ a new set of $p^{\pm}$ are determined with the help of equations (\ref{p+p-}). A new thin homogeneous layer is added at $x=L-2dx$ with the corresponding values of $\rho_e$ and $K_e$. Using the updated values of $p^{\pm}$ at $x=L$, the pressure subfields $p^{\pm}(L-dx,\omega)$ are calculated. The operation is repeated until the last infinitely thin layer is added at $x=0$. For each addition of a new layer, the continuity conditions on $p$ and $\phi(x) \mathbf{\dot{U}}\cdot\mathbf{n}$ are implicitely accounted for on both sides of the cumulated slab.

\emph{Green's function approach.---}The initialization of the procedure described above requires that $p^{\pm}$ must be determined at $x=L$. To avoid this calculation, a Green's function approach \cite{Sail,PIER} can be used. Two Green's functions $G^\pm$ are defined by 
\begin{equation}
	\forall x\in [0,L], \quad p^{\pm}(x,\omega)=G^{\pm}(x,\omega)p^+(L,\omega).
	\label{G+-}
\end{equation}
Green's functions are characteristic of the sole material properties and describe the internal field within the material. The boundary conditions at $x=L$ are known and are $G^+(L,\omega)=1$ and $G^-(L,\omega)=0$. The system of coupled first order linear differential equations in $G^{\pm}$ obtained by inserting (\ref{G+-}) in (\ref{p+p-}) can be solved numerically using a Runge-Kutta routine.

The reflection and transmission coefficients $R(\omega)$ and $T(\omega)$ are deduced from $p^{\pm}$ 
\begin{eqnarray}
    p^-(0,\omega) & = & R(\omega)p^+(0,\omega), \label{R} \\
    p^+(L,\omega) & = & T(\omega)p^+(0,\omega). \label{T}
\end{eqnarray}
From these coefficients, the reflected and transmitted waves can be simulated. 

In the numerical simulations, the function chosen to create the changes in physical properties is a distribution obtained from integrating the Gaussian normal distribution: $I(x) \! \! =\! C(1-erf(-(x-x_0)/r))$ where $C$ is a constant, $x_0$ the position of the jump and $r$ a steepness factor. The steeper is the jump the finer the stepping $dx$ must be for better accuracy. In the simulations, 400 points were chosen to discretize the total slab and $dx \! =17.1/400 \! = 0.0428 mm$. The value chosen for $r$ was $r = 0.1dx$. Values of $r$ less than this value had little effect on the computed results. Smoothing the jump by taking $r =10dx$ resulted in an important reduction of the signal reflected at the interface between the two layers.

\emph{Experimental results - comparison with predictions.---}The experimental principle is shown in Fig. \ref{fig:slab} where an airborne ultrasonic wave is generated and detected at normal or oblique incidence by specially designed (ULTRAN) air-coupled transducers in a frequency range between $150$ and $250 kHz$. The incident wave is partly reflected, partly transmitted and partly absorbed in the inhomogeneous porous layer. The materials studied are highly porous polyurethane foams saturated by air. The layers are put in contact, not glued. The physical parameters used for each layer were determined using a previous method \cite{Phil96} and are displayed in Table I.

\begin{figure}[hb]
\begin{minipage}{12.0cm}
		\centering\includegraphics[width=0.8 \textwidth]{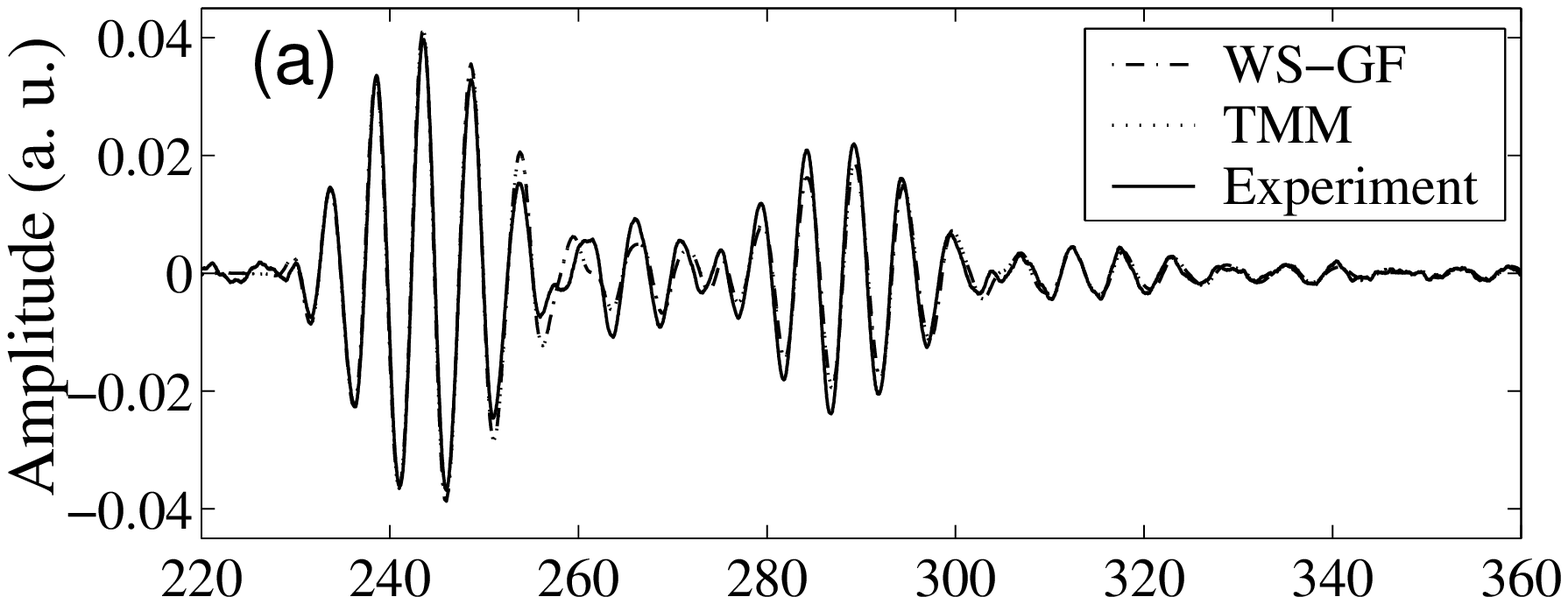}
\end{minipage}
\begin{minipage}{12.0cm}
		\centering\includegraphics[width=0.8 \textwidth]{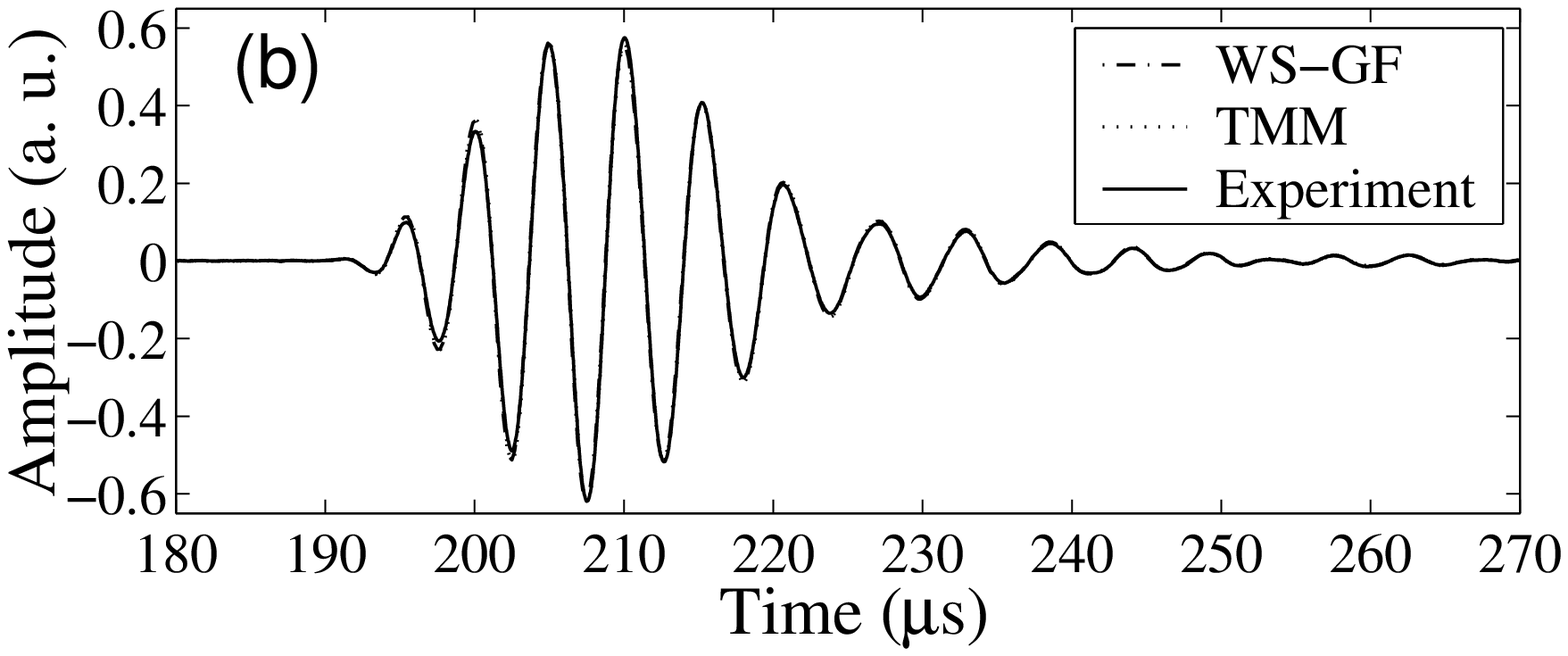}
\end{minipage}
	\caption{\emph{Experimental signal (a) reflected and (b) transmitted by a two-layer porous system. Comparison with the signals simulated by the WS-GF method and by the TMM method. The incidence is normal.}}
	\label{fig:transRT}
\end{figure}

\begin{figure}[ht]
	\centering
		\includegraphics[width=0.8 \textwidth]{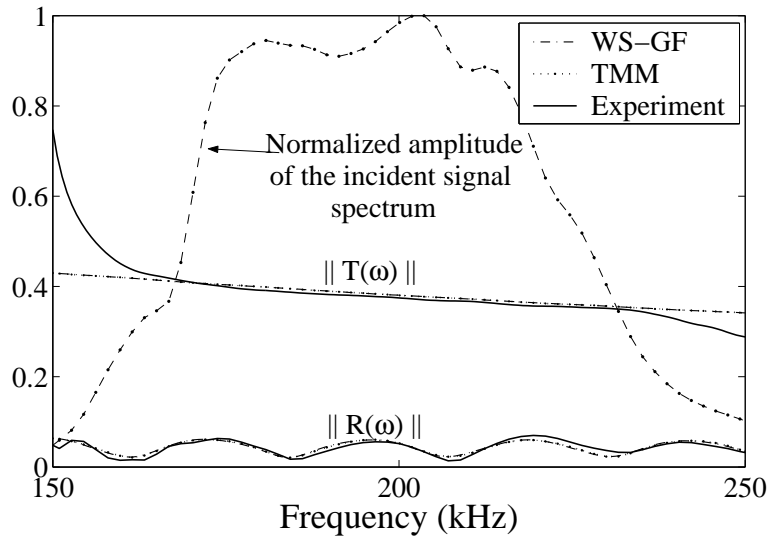}
	\caption{\emph{Modulus of the experimental normal incidence reflection and transmission coefficients $R(\omega)$ and $T(\omega)$. Comparison with the reflection and transmission coefficients simulated by the WS-GF method and by the TMM method.}}
	\label{fig:RT}
\end{figure}

\begin{table}[h]
\begin{tabular}{lccccccc}
\hhline{========}
 &  & $\phi$ & $\tau_{\infty}$ & $\Lambda$ & $\Lambda'$ & $R_f$ & Thickness \\
 &        &              &     & $(\mu m)$ & $(\mu m) $&$(Ns.m^{-4})$& $(mm)$ \\
\hline
 \textbf{Layer 1} &  & 0.96 	& 1.07 	& 273 & 672 & 2843  & 7.1  \\
 \textbf{Layer 2} &  & 0.99 & 1.001 & 230 & 250 & 12000 & 10.0\\
\hhline{========}
\end{tabular}
\label{tab1}
\caption{\emph{Properties of the two-layer medium studied.}}
\end{table}
\noindent The reflection experiment was carried out with a single transducer used in the pulse-echo mode \cite{Fellah2000}. Another transducer was required on the other side of the specimen to measure the transmission. A particularly good agreement between the experimental results and the results of the simulations is found for the reflected and transmitted signals (Fig. \ref{fig:transRT}(a) and \ref{fig:transRT}(b)). In Fig. \ref{fig:transRT}(b), the three curves are almost indistinguishable. The waveforms were calculated from the experimental incident waveform, which was recorded and introduced into the simulation routines. The agreement is also very good for the reflection and transmission coefficients (Fig. \ref{fig:RT}) in the frequency range $170 - 230 kHz$. In all figures, the WS-GF-curves cannot be distinguished from the TMM-curves. The discrepancies below $170 kHz$ and above $230 kHz$ in Fig. \ref{fig:RT} are attributed to the limited bandwidth of the incident signal, making the experimental results more sensitive to noise outside the useful bandwidth.
\clearpage

\emph{Conclusion.---}The equations of motion in fluid-saturated inhomogeneous porous media were derived from Biot's second formulation of 1962 \cite{BiotJAP62}. A wave equation, valid in the rigid frame approximation, was also proposed and solved for a two-layer system of homogeneous porous material considered as one single inhomogeneous porous layer with a sudden change in its physical properties. The transition from the properties of the first layer to those of the second layer was modeled by continuous inhomogeneity function with the shape of the Heaviside function. Excellent agreement was found between experimental and simulated waveforms and also between their respective reflection and transmission coefficients in a frequency range between $170$ and $230 kHz$. Future developments of this work are oblique incidence experiments and the study of other types of inhomogeneity profiles such as the linear profile. This work should contribute to the determination of the inhomogeneity profile of unknown materials by the use of inversion methods.

\emph{Acknowledgments.---}We are grateful to Prof. O. Matsuda for his useful comments on earlier versions of this paper.


\begin{thebibliography}{}
\bibitem{Aspnes} D. E. Aspnes and A. Frova, Solid State Comm., \textbf{7}, 155 (1969). 
\bibitem{Osamu} O. Matsuda and O. B. Wright, J. Opt. Soc. Am. B, \textbf{19}, 3028 (2002).  
\bibitem{PIER} J. Lunstedt and M. Norgren, Progress In Electromagnetics Research, PIER \textbf{43}, 1 (2003).
\bibitem{Wirgin} J. L. Buchanan, R. P. Gilbert, A. Wirgin and Y. S. Xu, \emph{Marine Acoustics: Direct and Inverse Problems}, (SIAM, Philadelphia, 2004).
\bibitem{PRL94} D. J. Van Manen, J. O. A. Robertsson, and A. Curtis, Phys. Rev. Lett. \textbf{94}, 164301 (2005).
\bibitem{Berryman} J. G. Berryman and R. R. Greene, Geophysics \textbf{45}, 213 (1980). 
\bibitem{TournatPRL} V. Tournat, V. Zaitsev, V. Gusev, V. Nazarov, P. B\'equin, and B. Castagn\`ede, Phys. Rev. Lett. \textbf{92}, 085502 (2004). 
V. Tournat, V. E. Gusev, and B. Castagn\`ede, Phys. Rev. E \textbf{70}, 056603 (2004).
\bibitem{Jia} X. Jia, Phys. Rev. Lett \textbf{93}, 154303 (2004).
\bibitem{Biot56} M. A. Biot, J. Acoust. Soc. Am., \textbf{28}, 168 (1956).
\bibitem{John87} D. L. Johnson, J. Koplik, and R. Dashen, J. Fluid Mech., \textbf{176}, 379 (1987).
\bibitem{Att1} K. Attenborough, Phys. Rep., \textbf{82}, 181 (1982).
\bibitem{Allard} J. F. Allard, \emph{Propagation of Sound in Porous Media: Modeling Sound Absorbing Materials}, (Chapman and Hall, London, 1993).
\bibitem{BiotJAP62} M. A. Biot, J. Appl. Phys., \textbf{33}, 1482 (1962).
\bibitem{Carcione} J. M. Carcione, \emph{Wavefield in real media: Wave propagation in anisotropic, anelastic and porous media}, (Pergamon, Amsterdam, 2001), Vol.31 p.261.
\bibitem{BiotWillis} M. A. Biot and D. G. Willis, J. Appl. Mech., 594 (1957).
\bibitem{Champ} Y. Champoux and J. F. Allard, J. Appl. Phys., \textbf{70}, 1975 (1991).
\bibitem{Sail} S. He, J. Math. Phys, \textbf{34}, 4628 (1993).
\bibitem{Phil96} P. Leclaire, L. Kelders, W. Lauriks, N. R. Brown, M. Melon, and B. Castagn\`ede, J. Appl. Phys., \textbf{80}, 2009 (1996).
\bibitem{Fellah2000} Z. E. A. Fellah, S. Berger, W. Lauriks, C. Depollier, P. Trompette, and J. Y. Chapelon, J. Appl. Phys., \textbf{93}, 9352 (2003).
\end{thebibliography}
\end{document}